\documentclass[a4paper, 10pt, conference]{ieeeconf}      
\IEEEoverridecommandlockouts                              
\usepackage[]{graphicx}
\usepackage{subfigure} 
\usepackage{amsmath} 
\usepackage{amssymb}  
\renewcommand{\theenumi}{(\alph{enumi}}
\graphicspath{{Figures/}}

\usepackage[noadjust]{cite}

\usepackage[implicit=false,hidelinks]{hyperref}
\usepackage{url}
\usepackage{balance}

\newcommand{\x}{\mathbf{x}}
\newcommand{\X}{\mathbf{X}}
\newcommand{\z}{\mathbf{z}}
\newcommand{\h}{\mathbf{h}}

\newcommand{\Z}{\mathbf{Z}}

\newcommand{\K}{\mathbf{K}}

\newcommand{\I}{\mathbf{I}}

\newcommand{\lat}{\mathbf{l}}

\newcommand{\var}{\mathrm{var}}

\newcommand{\E}{\mathbb{E}}

\title{Distance Invariant Sparse Autoencoder for\\Wireless Signal Strength Mapping}
\author{Renato Miyagusuku and Koichi Ozaki
\thanks{*This work was supported by the National Institute of Information and Communications Technology (NICT)}
\thanks{R. Miyagusuku and K. Ozaki are with the Department of Mechanical and Intelligent Engineering, Utsunomiya University, Japan.
{\footnotesize \{miyagusuku, ozaki\}@cc.utsunomiya-u.ac.jp}}}

\usepackage{fancyhdr}
\usepackage{ragged2e}

\begin{document}
\maketitle

\thispagestyle{fancy}
\renewcommand{\headrulewidth}{0pt} 
\chead{\justify
 \vspace*{-3cm} \scriptsize {\textcopyright 2020 IEEE. Personal use of this material is permitted. Permission from IEEE must be obtained for all other uses, in any current or future media, including reprinting/republishing this material for advertising or promotional purposes, creating new collective works, for resale or redistribution to servers or lists, or reuse of any copyrighted component of this work in other works. 
}}
\pagestyle{empty}


\begin{abstract}
Wireless signal strength based localization can enable robust localization for robots using inexpensive sensors. For this, a location-to-signal-strength map has to be learned for each access point in the environment. Due to the ubiquity of Wireless networks in most environments, this can result in tens or hundreds of maps. To reduce the dimensionality of this problem, we employ autoencoders, which are a popular unsupervised approach for feature extraction and data compression. In particular, we propose the use of sparse autoencoders that learn latent spaces that preserve the relative distance between inputs. Distance invariance between input and latent spaces allows our system to successfully learn compact representations that allow precise data reconstruction but also have a low impact on localization performance when using maps from the latent space rather than the input space. We demonstrate the feasibility of our approach by performing experiments in outdoor environments.

\end{abstract}
\section{Introduction}
Wireless signal strength based localization has been successfully used for indoor and outdoor robot localization, achieving around 1~m accuracy indoors without the need to modify the environment or acquiring extensive training datasets~\cite{miyagusuku2019data}, and around 3-5~m outdoors~\cite{benjamin2015real}. Compared to localization systems that use cameras~\cite{chancan2020hybrid} or range data~\cite{hess2016real}, the accuracy of wireless signal strength based localization systems is low; however, wireless signal strength localization systems possess other appealing characteristics: wireless signals do not suffer from the data association problem, do not require the installation of expensive hardware, and require relatively low computation~\cite{miyagusuku2018precise}.

A common approach to designing localization systems using wireless signal strength measurements is to learn location-to-signal strength mappings for each access point in the area of interest and use these mappings to compute the likelihood of a location given new measurements using Bayes-filters. In particular, we use Gaussian Processes (GPs) for learning these mappings, as their effectiveness has been demonstrated in practice~\cite{ferris2006gaussian,miyagusuku2016improving,zhao2018gaussian,wang2020indoor}. GPs are a generalization of normal distributions to functions. They generalize a finite amount of training data pairs into a continuous function, where each point has a normal distribution. 

Considering that office building can have several tens to a few hundreds of access points, we need to potentially use hundreds of maps. While an abundance of data is beneficial to improve localization accuracy and robustness, it also increases the required storage and computational resources. This problem is exacerbated for outdoor localization~\cite{miyagusuku2020toward}, as the number of access points considerably increases as well as the map area required which can become a problem for resource-constraint edge devices.

As wireless signal strength maps are highly correlated~\cite{miyagusuku2019data}, more compact representations can be learned without causing significant data degradation; alleviating the aforementioned problems. Several approaches have been presented in the literature for dimensionality reduction, such as principal component analysis, and autoencoders. In this work, we propose a novel approach that learns a lower-dimensional latent space using sparse autoencoders that encourages preserving the relative distance between data points in the input space (signal strengths) and in the compressed latent space.

\begin{figure}[bt]
 \centering
 \vspace{10 pt}
 \includegraphics[width=\columnwidth]{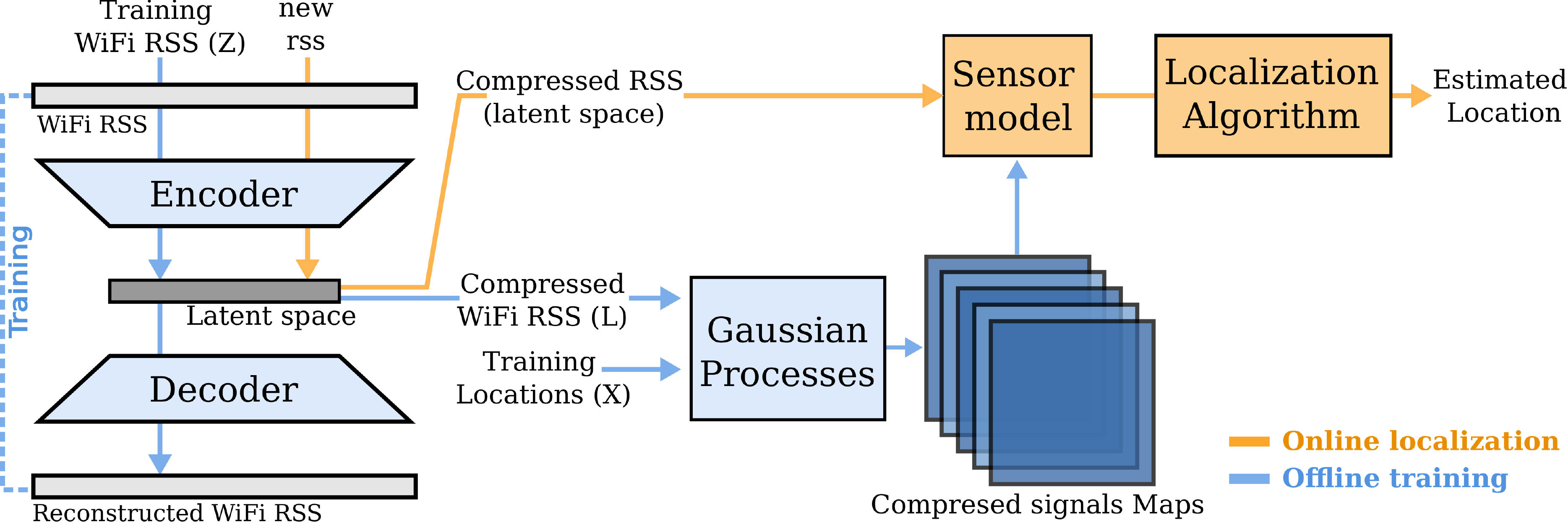}
 \caption{Overview of our approach. Using our proposed distance invariant autoencoder, we learn a compact representations of RSS data, which can be used to learn signal strength maps and be used directly for Localization. Our proposed system requires less memory for storage and allows faster localization with almost equal accuracy. \label{fig:diagram}}
\end{figure}  

By enforcing distance invariance we preserve most properties of the input space which allows our GPs formulation to perform equally well. This results in mappings of the latent space which can be used as sensor models with little impact on the quality of the posteriors generated. In this work, we propose learning such latent spaces, so we can generate maps in the compressed space and use them directly for wireless-based localization (Fig.~\ref{fig:diagram} shows our overall approach).

The remaining of this work is organized as follows. In section \ref{sec:related} we discuss related works in wireless signals-based localization, as well as autoencoders and their application to this field. In section \ref{sec:approach} we introduce our approach as well as the reasoning and justifications for our distance invariance condition. In section \ref{sec:experiment} we present experimental data acquired in outdoor scenarios, which demonstrate the feasibility of our approach. Finally, section \ref{sec:conclusion} presents this work's conclusions and future works.

\section{Related Work\label{sec:related}}
\subsection{Localization using wireless signals}
Two signal metrics are commonly used for wireless signals-based localization: Received Signal Strength (RSS) information and Channel State Information (CSI). RSS refers to the signal strength in dBm a signal has at a given location and has been adopted by the majority of wireless-based localization approaches for its ease of use. There is no need for special hardware like antenna arrays, nor modified firmware. While path loss models can be used to estimate the distance from signal strength measurements, signal strength is a poor estimator of distance as wireless signals propagation through space is affected by its reflection and refraction with all surrounding objects and is characterized by complex shadowing and multipath effects - which are not considered in most path loss models~\cite{benkivc2008using}. CSI matrices represent information based on the state of a communication link and require at least two receivers and transceiver antennas. CSI is more robust to the ill effects of multipath and, in general, can be used to achieve higher localization accuracies~\cite{yang2013rssi}. However, its use requires the use of modified firmware such as the one presented in \cite{halperin2011tool,xie2015precise}. While this work employs RSS information, the developed system can be easily adapted for its use with CSI data.

To use any of these metrics for localization, Fingerprinting techniques are the most widely used. Fingerprinting refers to techniques in which given measurement samples are acquired at known locations in the environment, an algorithm is trained to predict the location of new measurements. Fingerprinting can be done by matching new measurements to the most similar samples in the training dataset, which we classify as direct methods as inputs are wireless signals, outputs are locations; or by learning a location-to-signal mapping from the training dataset, and then computing the likelihood of new measurements to have been originated at any candidate location, which we classify as indirect methods. Examples of direct methods used for wireless signals-based localization include the use of random forest~\cite{benjamin2015real}, support vector machines~\cite{del2011comparison}, convolutional~\cite{song2019novel} and recurrent~\cite{hsieh2018towards} neural networks. Examples of indirect methods include learning graphs and performing linear interpolation in graphs~\cite{biswas2010wifi}, vector field maps~\cite{gutmann2012vector}, or signal strength maps using Gaussian Processes~\cite{ferris2006gaussian,miyagusuku2016improving,zhao2018gaussian,wang2020indoor}.

The main advantage of GPs over direct and other indirect methods is that GPs mathematical formulation derives predicted variances directly; i.e., we can make mean and variance predictions. This allows us to easily incorporate GP predictions into sensor models and well studied probabilistic models for localization such as Monte Carlo Localization or any other Bayes-filter.

\subsection{Dimensionality reduction}
Dimensional reduction is the problem of learning a transformation from a higher-dimensional input space, to a latent space while preserving as much as possible of the variation of the input space. Highly correlated input spaces can be compressed to low-dimensional latent spaces without much information loss.

Several approaches have been proposed for dimensional reduction. A classic approach is principal component analysis (PCA). In PCA, given a dataset sampled from the input space of dimension $m$, PCA computes the dataset's matrix eigenvectors $\nu$ and sorts them in decreasing corresponding eigenvalues $\lambda$. For a latent space of dimension $c$, the first $c$ eigenvectors are selected to form a $m\times c$ transformation matrix. The $c$ eigenvectors with the largest eigenvalues are selected as those explain more of the variation in the data.

Another approach that has become popular in recent years due to the advent of deep learning is the use of autoencoders. Autoencoders are neural networks trained to reproduce their input as to their output. Formally, given an input vector $\z$, the network learns a function $f$, often referred as \textit{encoder}, that maps it to a vector $\h$ (${f: \z\rightarrow\h}$), and a function $g$, often referred as \textit{decoder}, that maps this vector to a vector $\hat{\z}$ ((${g: \h\rightarrow\hat{\z}}$)) that should reproduce the input vector as closely as possible, i.e., ${\hat{\z} \approx \z}$. Therefore, autoencoders are trained using a loss function that minimizes the difference between its inputs and outputs. Opposed to other PCA, autoencoders are usually nonlinear and can learn more complex latent spaces - though the possibility of overfitting needs to take into consideration, especially if the encoder/decoder used is a deep network.

Autoencoders whose intermediate vectors' dimension is lower than that of its inputs are referred to as undercomplete autoencoders. Lower dimensionality on the intermediate vector forces the autoencoder to learn compact representations and have been successfully used for dimensional reduction, feature extraction, clustering, etc. Autoencoders whose intermediate vectors' dimension is larger than that of its inputs, are referred to as overcomplete autoencoders and have been used to improve robustness and as generative models.

Some variants of autoencoders that have been previously proposed include sparse, contractive, and denoising autoencoders. Sparse autoencoders add a sparsity penalty on the intermediate vector $\Omega(\h)$ to the loss function used for training. This penalty encourages learning models were few nodes in the intermediate layer are active at the same time, achieving sparse models~\cite{goodfellow2009measuring}. Contractive autoencoders add the penalty to the derivative of the intermediate vector, which encourages locally consistent models.  Small perturbations on the input space result in similar intermediate vectors~\cite{rifai2011higher}. Denoising autoencoders add noise to its inputs during training, which encourages learning more robust models~\cite{vincent2008extracting}.

Autoencoders have been used for a wide range of applications. For wireless signals-based localization, autoencoders have been used to learn latent spaces from RSS data, which were then used for classification using either a fully connected neural network~\cite{chidlovskii2019semi} or a convolutional neural network~\cite{song2019novel}. As well generative models for CSI data augmentation~\cite{chen2020fido}. 

In this work, we propose the use of autoencoders to learn and then use the complete latent manifold instead of single data points (as in previous works). For this, we are also interested in the latent space to retain certain characteristics of the input space's structure. We detail our approach in the following section.

\section{Autoencoder for Wireless Signal Strength Mapping\label{sec:approach}}
Our overall approach (shown in Fig.~\ref{fig:diagram}), uses an autoencoder to learn compact representations of the signals in one environment so that signal strength maps in this latent space can be used directly for wireless signals-based localization. 

For this, we propose a new autoencoder which preserves the relative distance between two points in the input space and its latent space (intermediate layer with lower dimension). Once the autoencoder has been learned, training data is transformed into this latent space using its encoder. Using a GP we learn location-to-signal mappings in this latent space and use the resulting maps directly for robot localization. 

Distance invariance between input and latent spaces allows the mappings generated using our GP model to retain the same properties on either space. In the remainder of this section, we first describe our approach for wireless signal strength mapping using Gaussian Processes, to then delve into our distance invariant sparse autoencoder. 

\subsection{Mapping using Gaussian Processes}
Formally, given some training data $(\X,\Z)$ where $\X \in \mathbb{R}^{n\times 2}$ is the matrix of $n$ input samples locations (x-y coordinates) $\x_i, \in \mathbb{R}^2$; and $\Z \in \mathbb{R}^{n\times m}$ the matrix of corresponding output samples $\z_{i} \in \mathbb{R}^m$. Two assumptions are made. 

First, each data pair $(\x_i,\z_i)$ is assumed to be drawn from a noisy process: 
\begin{equation}
\z_i = f(\x_i)+\epsilon,
\label{eq:iid_assumption}
\end{equation} 
where $\epsilon$ is the noise generated from an identical Gaussian distribution with known variance $\sigma^2_n$ - though this assumption can be relaxed using heteroscedastic Gaussian Processes~\cite{miyagusuku2015gaussian}. 

Second, any two output values, $\z_p$ and $\z_q$, are assumed to be correlated by a covariance function based on their input values $\x_p$ and $\x_q$:
\begin{equation}
cov(\z_p,\z_q)=k(\x_p,\x_q)+\sigma^2_n\delta_{pq},
\label{eq:standard_gp_covariance}
\end{equation} 
where $k(\x_p,\x_q)$ is a kernel, $\sigma^2_n$ the variance of $\epsilon$ and $\delta_{pq}$ is one only if $p=q$ and zero otherwise.

Given these assumptions, for any finite number of data points, the GP $\mathcal{G}$ can be considered to have a multivariate Gaussian distribution:
\begin{equation}
\mathcal{G}: \z\ \sim\ \mathcal{N}(m(\x),\,k(\x_p,\x_q)+\sigma^2_n\delta_{pq}).
\label{eq:multiGaussian}
\end{equation} 

Assuming without loss of generality that the mean to be the zero function, i.e., $m(\x) = \mathbf{0}$, a GP its fully defined by its kernel function $k(\x_p,\x_q)$. 

The most commonly used kernel function is the squared exponential kernel, also commonly referred to as the radial basis function or the Gaussian kernel~\cite{williams2006gaussian}. This kernel is defined as:
\begin{equation}
 k_{rbf}(\x_p,\x_q) = \sigma_s^2\exp\left(-\frac{|\x_p-\x_q|^2}{l^2}\right),
 \label{eq:kern_rbf}
\end{equation}
with hyper-parameters $\sigma_s^2$ (known as the signal variance) and $l$ (known as the length-scale). This kernel is isotropic, i.e., it is invariant to any rigid motion. Invariance to rigid motion is the desired quality as it allows the space of $\x$ to be rotated and translated without requiring to re-calculate the kernel. Additionally, by correctly rescaling the length parameter, $\x$ can even be scaled without requiring to learn the kernel parameters again - though, recalculation of the kernel would be required.  

Predictions $\z_*$ for an unknown data point $\x_*$, can be done by conditioning $\z_*$ to $\x_*, \X$ and $\Z$, obtaining:
\begin{equation}
p(\z_*|\x_*,\X,\Z)\ \sim\ \mathcal{N}(\E[\z_*],\var(\z_*)),
\label{eq:fstardistribution}
\end{equation}
where,
\begin{eqnarray}
\E[\z_*] = \mathbf{k}_*^T(\K+\sigma^2_n\I_n)^{-1}\Z, \label{eq:standard_gpr_mean}\\
\var[\z_*] = k_{**}-\mathbf{k}_*^T(\K+\sigma^2_n\I_n)^{-1}\mathbf{k}_*, \label{eq:standard_gpr_var}
\end{eqnarray}
with $\K = cov(\X,\X)$ being the ${n\times n}$ covariance matrix between all training points $\X$, usually called Gram Matrix; $\mathbf{k}_*$ = $cov(\X,\x_*)$ the covariance vector that relates the training points $\X$ and the test point $\x_*$; and $k_{**}$ = $cov(\x_*,\x_*)$ the variance of the test point. As both the Gram Matrix as well as the training samples $\Z$ are constant, we can reformulate eq. (\ref{eq:standard_gpr_mean}) as,
\begin{equation}
 \E[\z_*] = \mathbf{k}_*^T\mathbf{W}, \label{eq:mean},
\end{equation}
with $\mathbf{W} \in \mathbb{R}^{n\times m}$ a constant matrix, becoming a linear function in the feature space $\Phi(\x):k(\x,\X)$. 

\subsection{Distance invariant Autoencoder}

\begin{figure}[bt]
 \centering
 \includegraphics[width=\columnwidth]{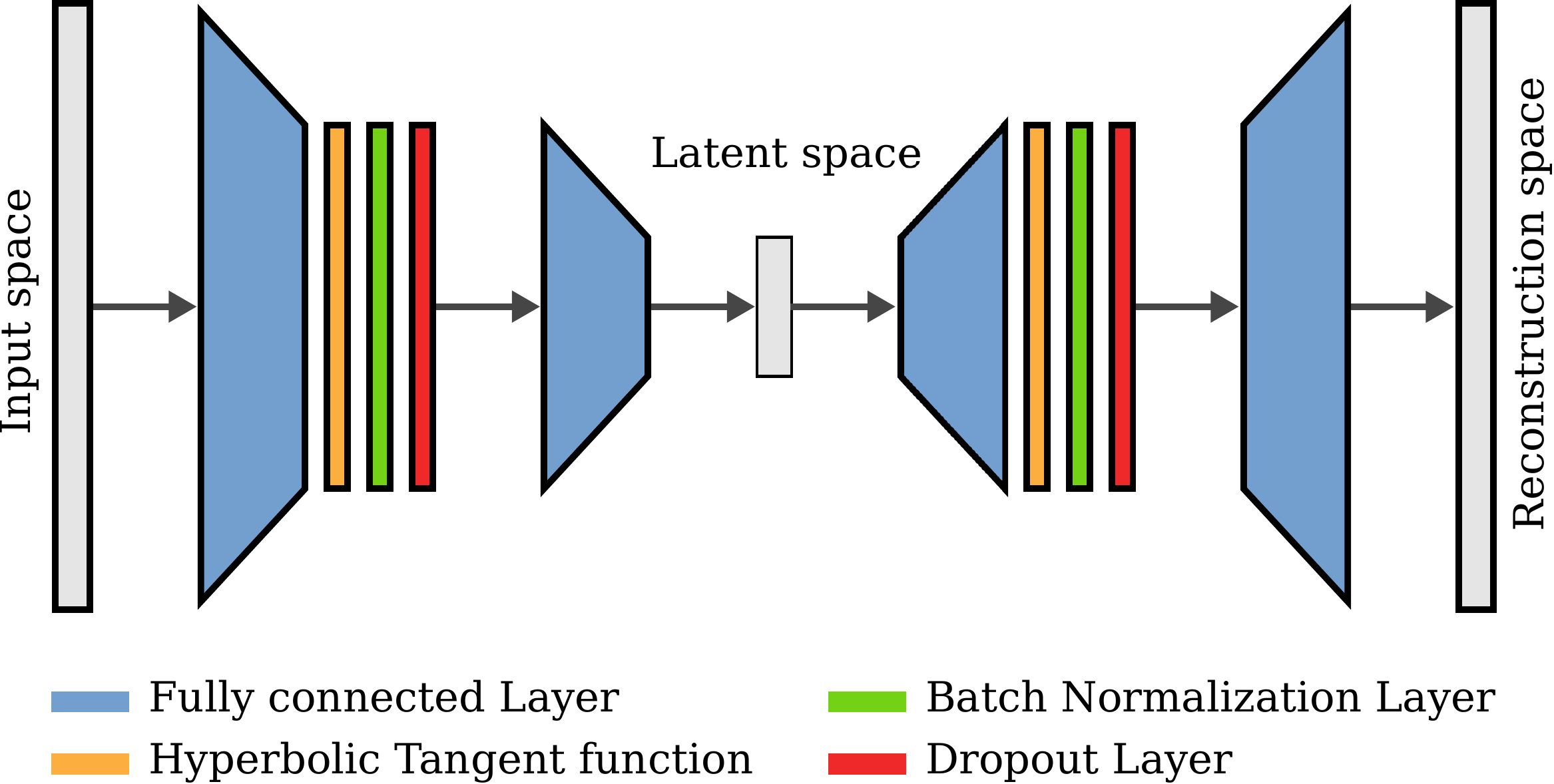}
 \caption{Proposed autoencoder structure\label{fig:autoencoder}}
\end{figure}  

In our approach, we model our encoder $f(\cdot)$ and decoder $g(\cdot)$ functions using two-layer fully connected neural networks, with the hyperbolic tangent as its activation function and batch normalization and dropout layers after the first layer. Figure \ref{fig:autoencoder} illustrates our model.

Shallow networks are used instead of deep networks due to the restricted amount of training data we have. While this approach is unsupervised, hence there is no need for manually tagging labels, the networks need to be trained with data specific to the environment where they will be deployed and acquisition of datasets in real environments using robots is time-consuming. Networks can not be generalized to work on all environments as the encoder exploits the correlations between the signals in the environment to learn more compact representations, hence intrinsically environment-specific. 

Nonetheless, we experimentally tested deeper networks. These deeper networks allowed us to learn more complex mappings and reconstruct the input space almost perfectly; however, this was most likely be a result of overfitting due to excessive capacity of networks as the latent functions we obtained did not yield better maps. 

\subsubsection{Training losses}
For training, we use the same sample matrix $\Z \in \mathbb{R}^{n\times m}$ and three loss functions.

A squared L2 norm as the reconstruction loss so $\hat{\z}\rightarrow\z$, 
\begin{equation}
    \mathcal{L}_r(\z,\hat{\z}) = \sum_{i\in b} |\z_i-\hat{\z_i}|^2
\end{equation}
where $b$ is a subset of the sample matrix (minibatch). A L1 regularization term on the latent space to encourage sparse outputs,
\begin{equation}
   \mathcal{L}_r(\lat) = \lambda_r\sum_{i \in b} |\lat_i|
\end{equation}
with $\lambda_r$ being a positive scalar. And a new training loss that we named distance invariance loss. This loss encourages isometry between input and latent spaces and is added to create a latent space that can be as easily reconstructed using GPs as the input space.

As shown in the previous section, $\E[\z]$ can be expressed as a linear function in the feature space $\Phi(\x):k(\x,\X)$. 
It is important to note that the feature space $\Phi(\x)$ depends on the family of kernel functions and the location dataset $\X$, both of which remain the same when learning using the input or latent spaces. To learn a latent space $\lat$ that retains as many as the properties as the GP learned using $\z$, a sensitive choice is to encourage $\lat$ to have the same structure as $\z$. A way to encourage that is by making both spaces isometric.

To encourage isometry of the spaces, we propose a loss function based on distance invariance. Specifically, we define the distance invariance loss as, 
\begin{equation}
    \mathcal{L}_d = \lambda_d\sum_{i \in b} \sum_{j \in b} |\z_i-\z_j|^2 - |\lat_i-\lat_j|^2
\end{equation}
with $\lambda_d$ being a positive scalar.

This distance invariance loss is a squared L2 norm on the difference between distances of the training points in the input and latent space, and while we use it to promote isometry between input and output spaces, it is important to note that we are only doing so for the training points, not the whole space. While our experiments show that this is sufficient for our case, further testing and research is required for isometry on the whole space.

\subsection{Maps and likelihoods in the latent space}
Finally, using the learned encoder $f$, we transform the training sample matrix $\Z$ to the latent sample matrix  ${\mathbf{L} = f(\Z)  \in \mathbb{R}^{n\times c}}$, and use it to create the new training dataset $(\X,\mathbf{L})$ for learning a GP map using the method described in the previous section. 

Using the encoder and this GP map, for any new measurement $\z$, we can compute the likelihood of an arbitrary location $\x_*$ to have originated the signal as
\begin{equation}
  p(\z|\x_*) = \prod_{d=0}^{c-1} \frac{1}{\var[\lat_{*}]^{0.5}}\varphi\left(\frac{\E[\lat_{*}]_d-f(\z)_d}{\var[\lat_{*}]^{0.5}}\right).
  \label{eq:likelihood}
\end{equation}

\begin{figure}[tb]
 \centering
 \includegraphics[width=\columnwidth]{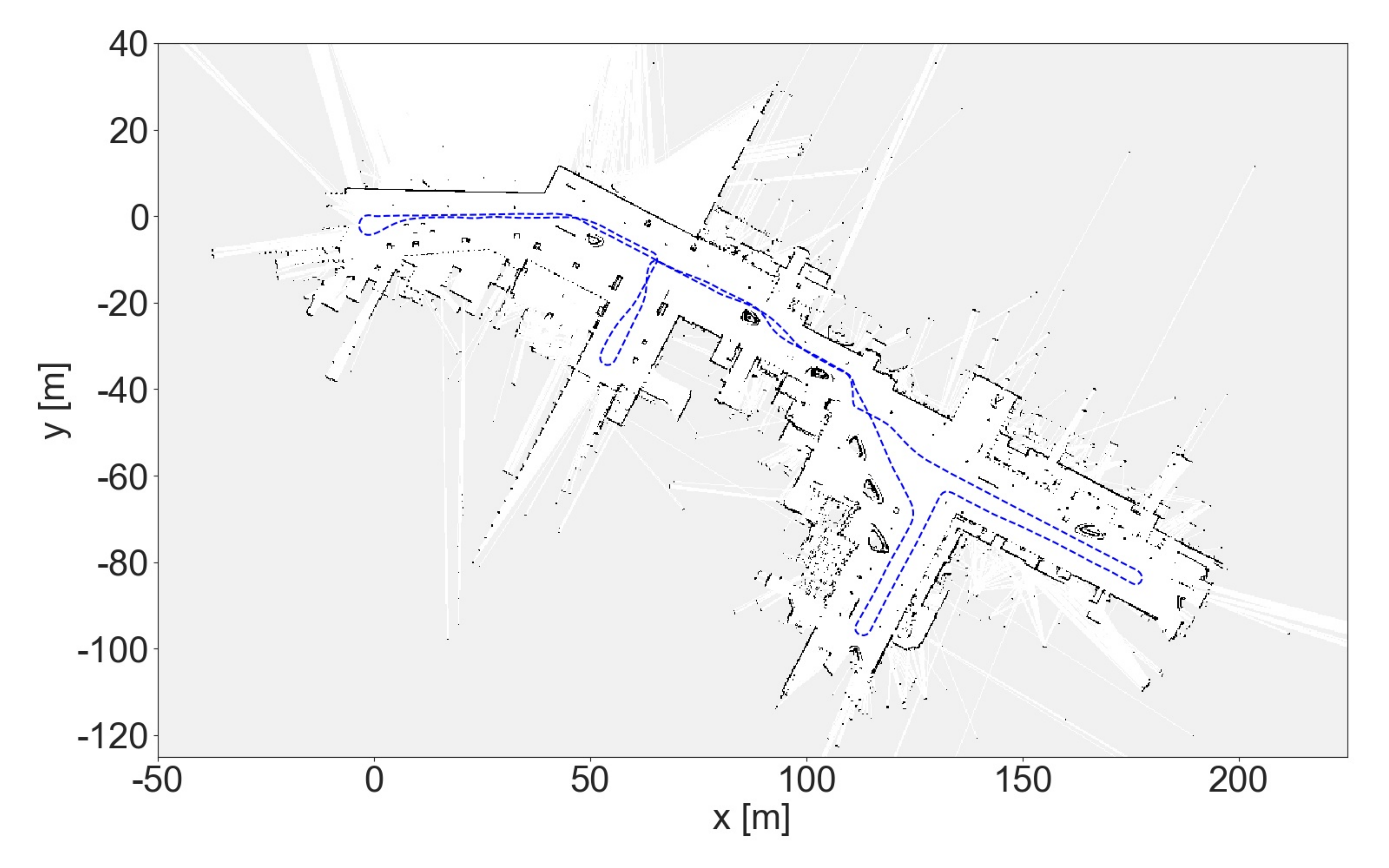}
 \caption{Occupancy grid map of the environment used for testing.\label{fig:hic_map}}
\end{figure}  

\section{Experimental evaluation\label{sec:experiment}}
\begin{figure*}[tb] 
  \centering 
  \raisebox{\dimexpr 1.25cm-\height}{(A)}~
  \subfigure{\includegraphics[width=.22\textwidth]{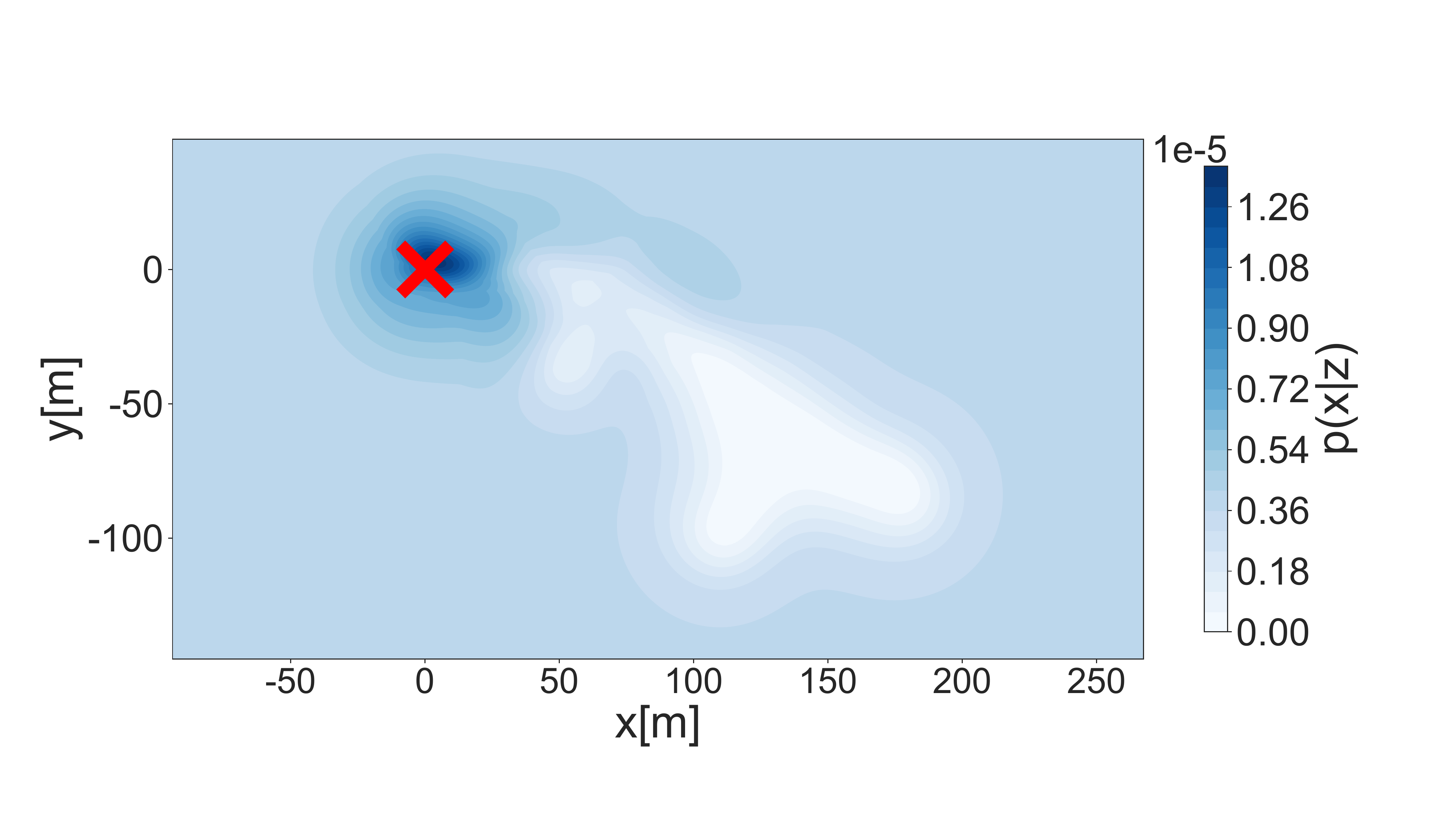}}~ 
  \subfigure{\includegraphics[width=.22\textwidth]{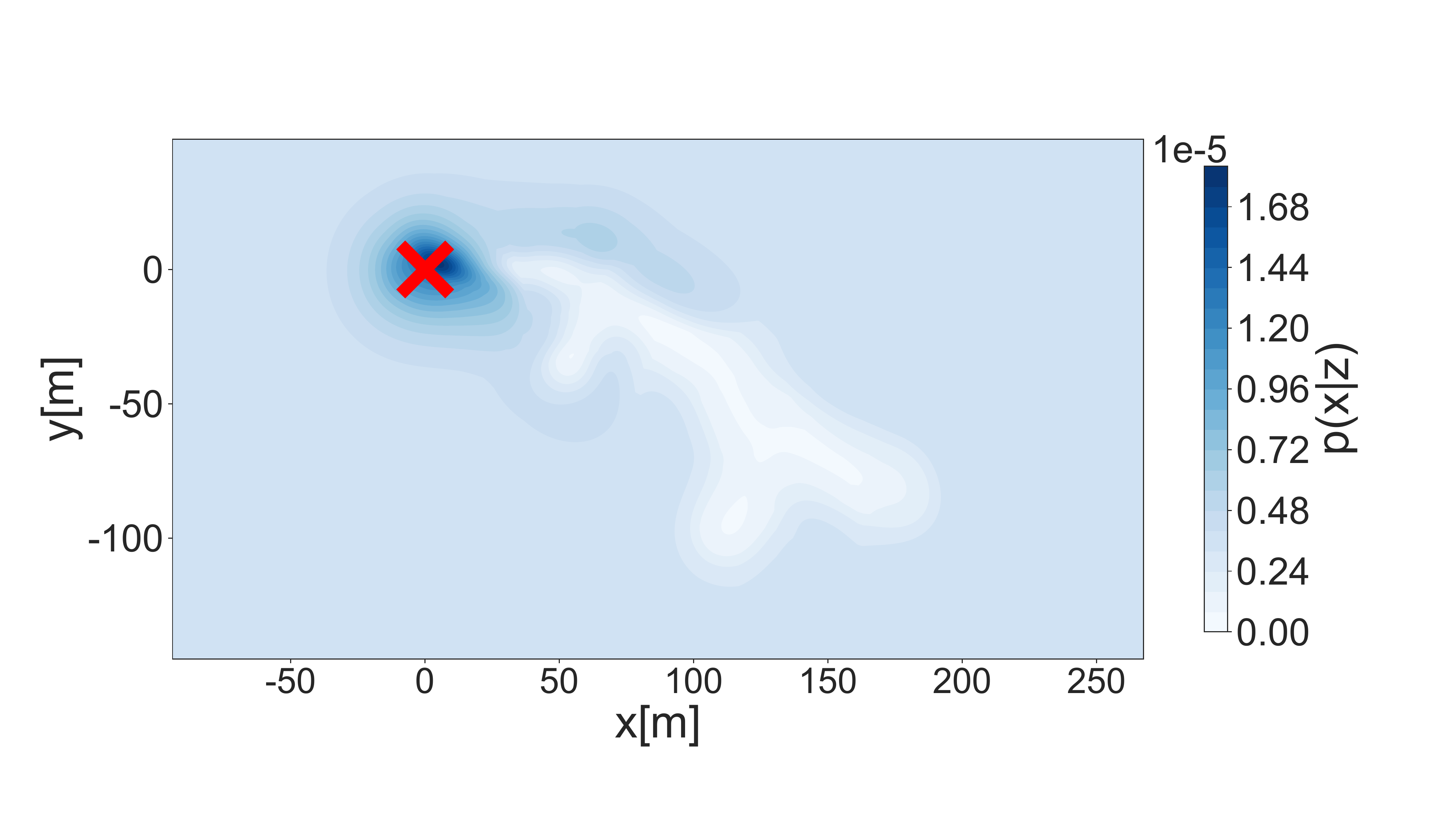}}~ 
  \subfigure{\includegraphics[width=.22\textwidth]{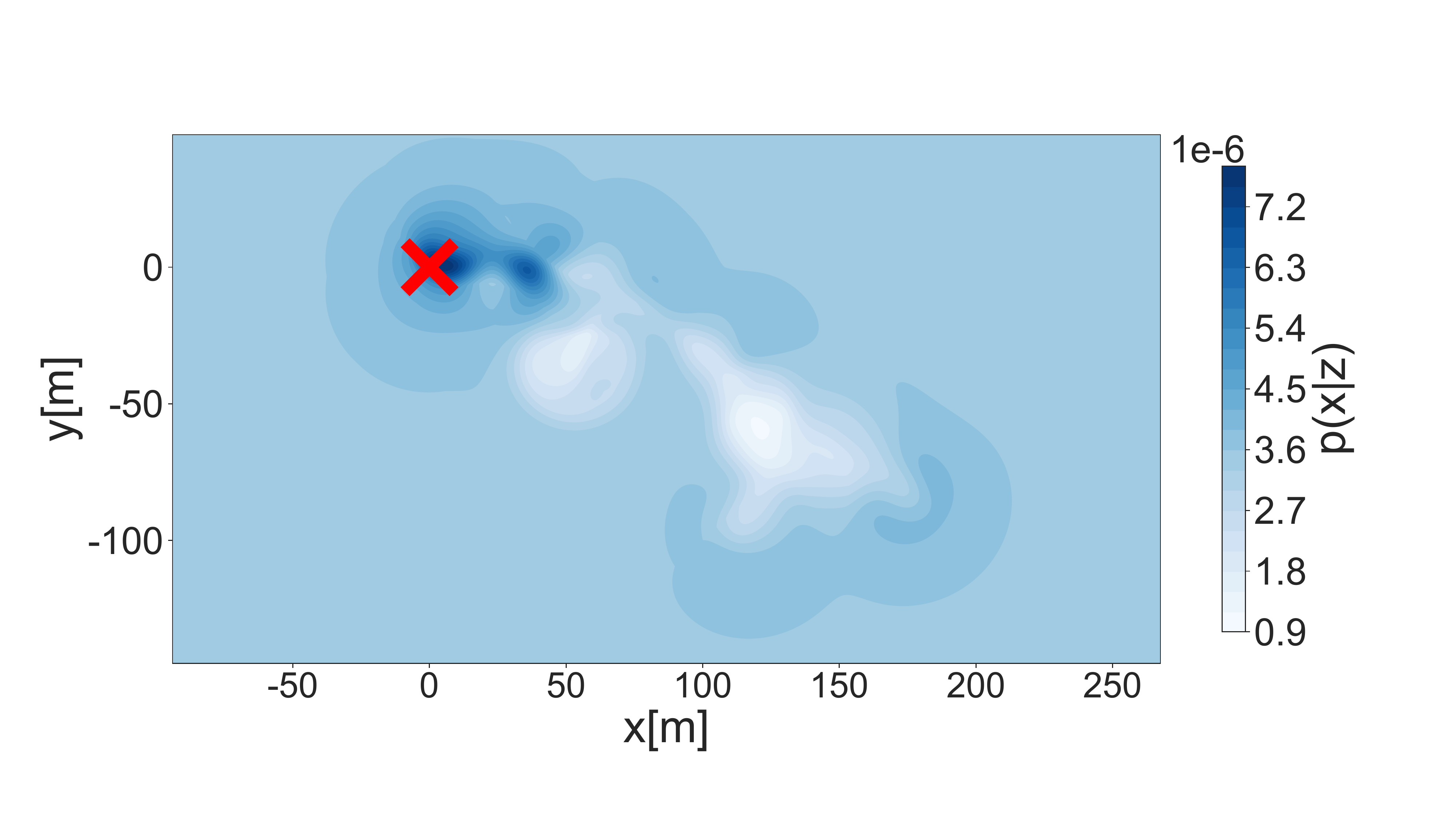}}~
  \subfigure{\includegraphics[width=.22\textwidth]{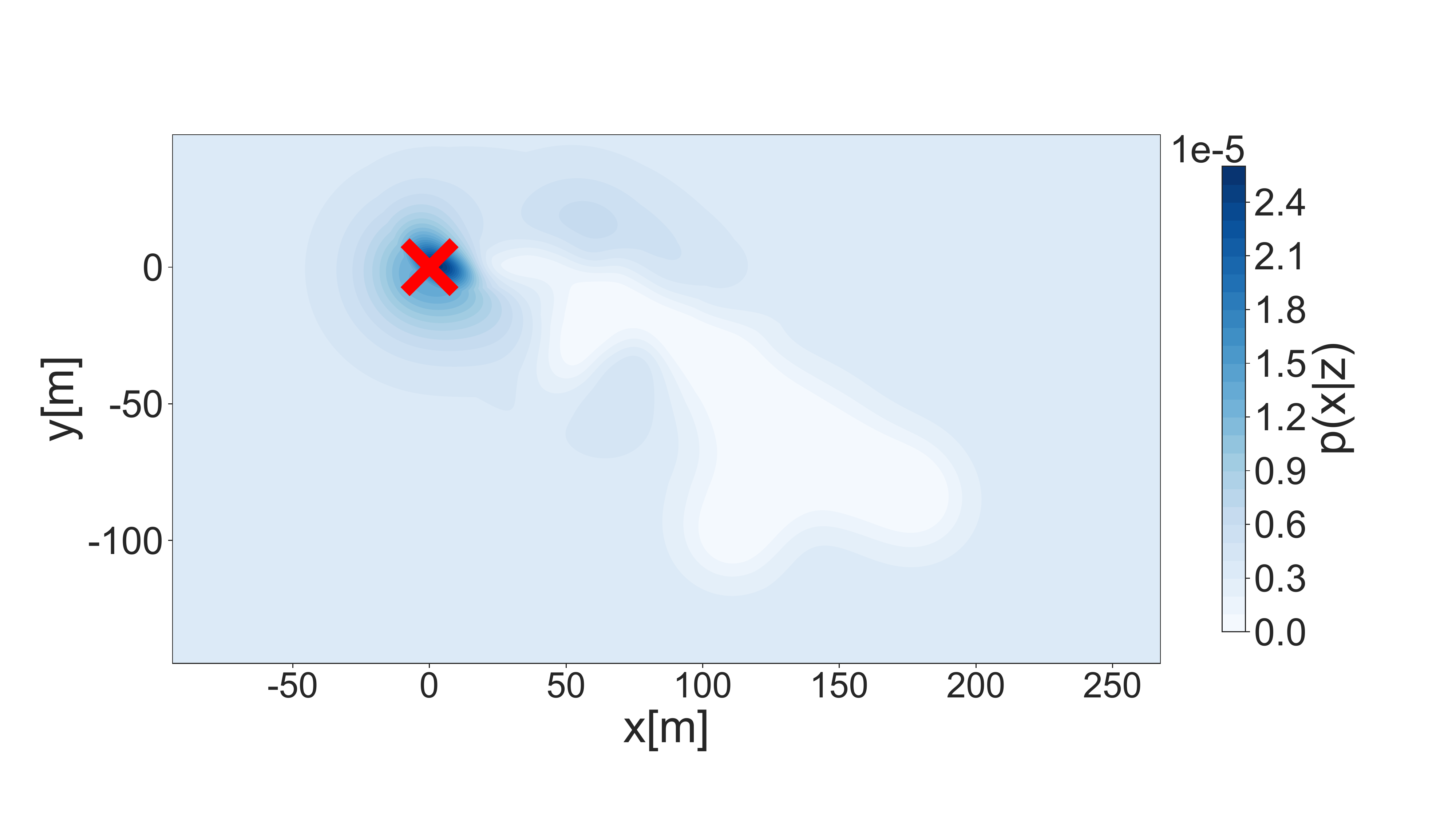}}\newline
  \raisebox{\dimexpr 1.25cm-\height}{(B)}~
  \subfigure{\includegraphics[width=.22\textwidth]{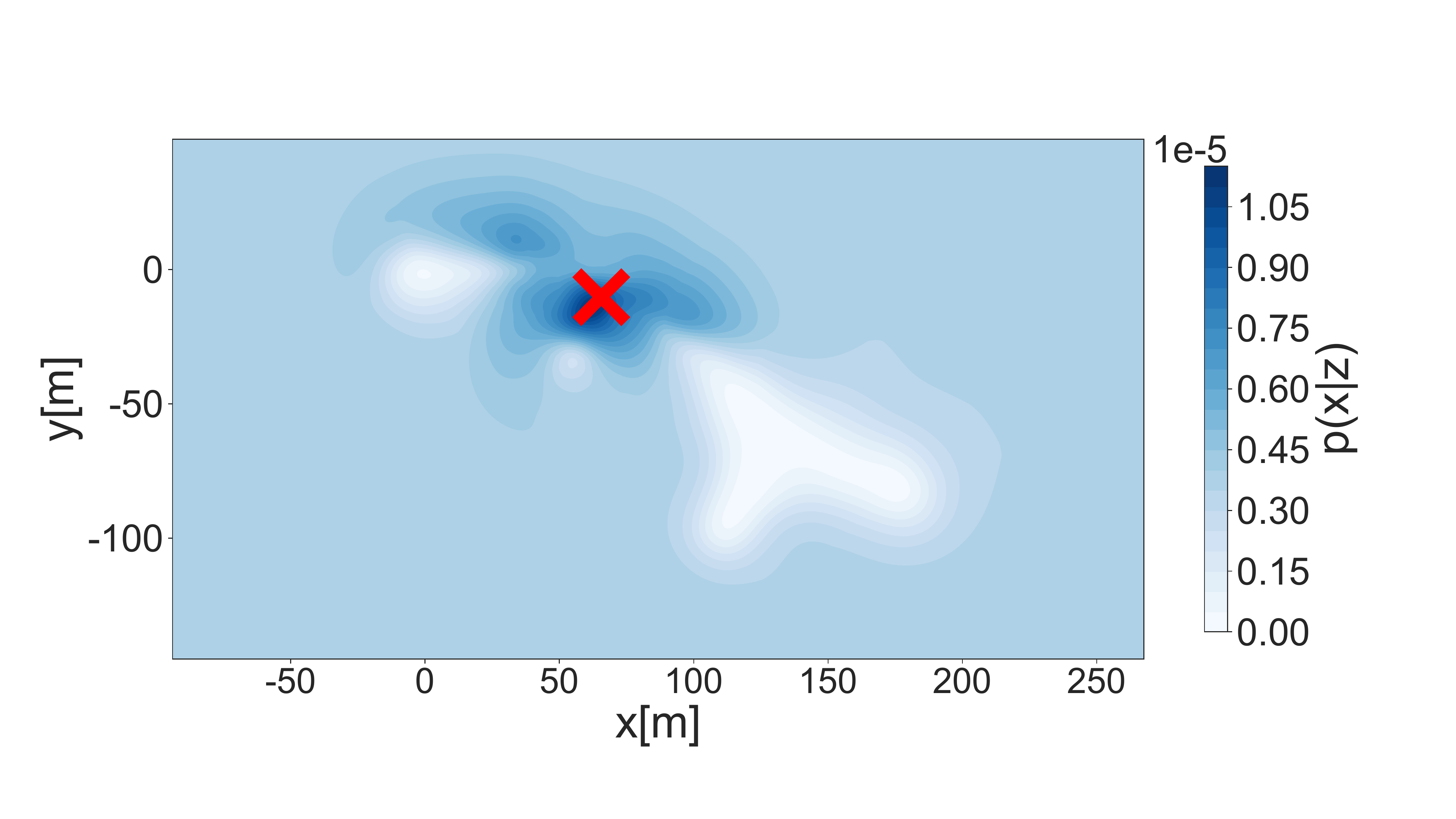}}~ 
  \subfigure{\includegraphics[width=.22\textwidth]{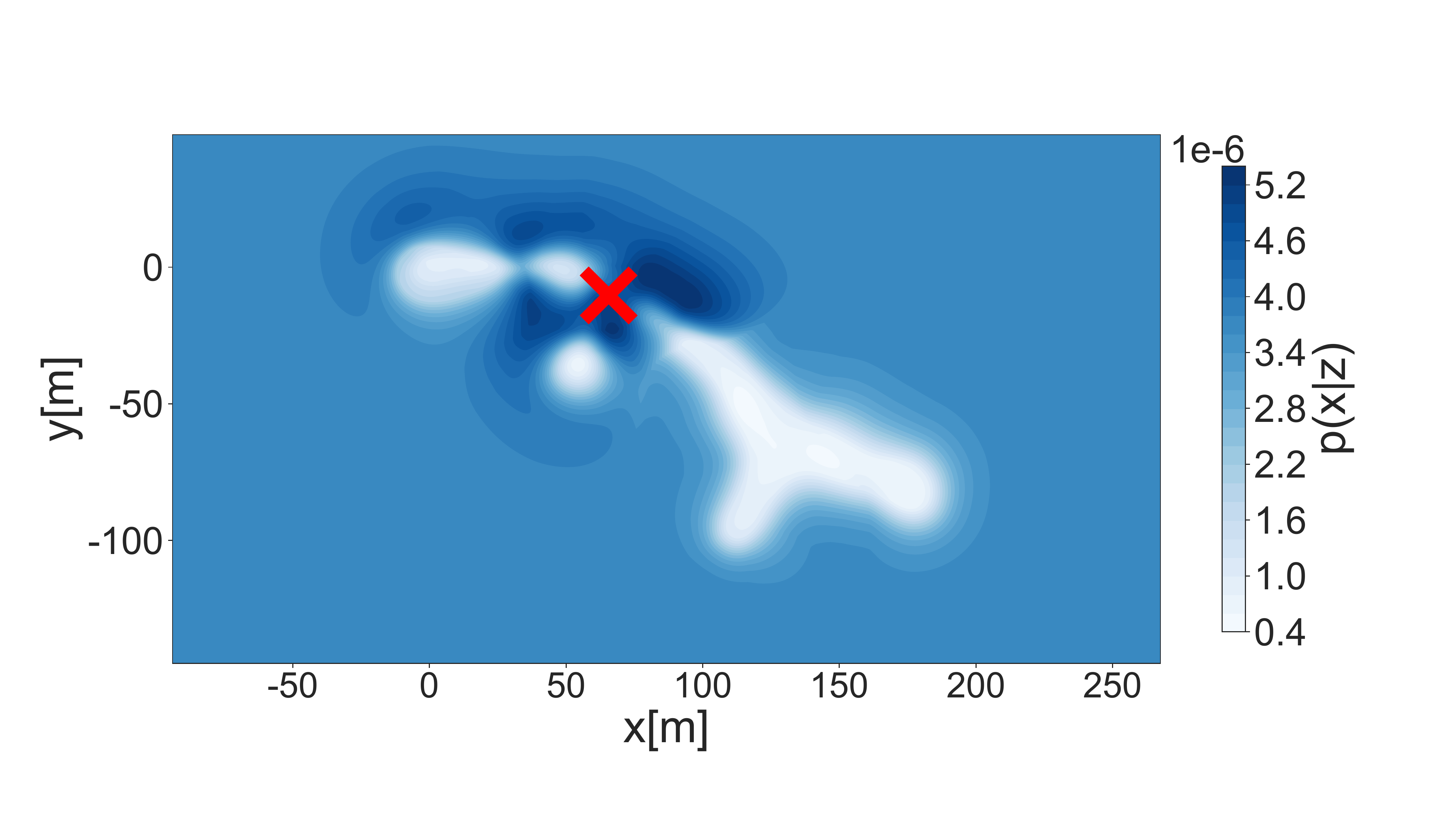}}~ 
  \subfigure{\includegraphics[width=.22\textwidth]{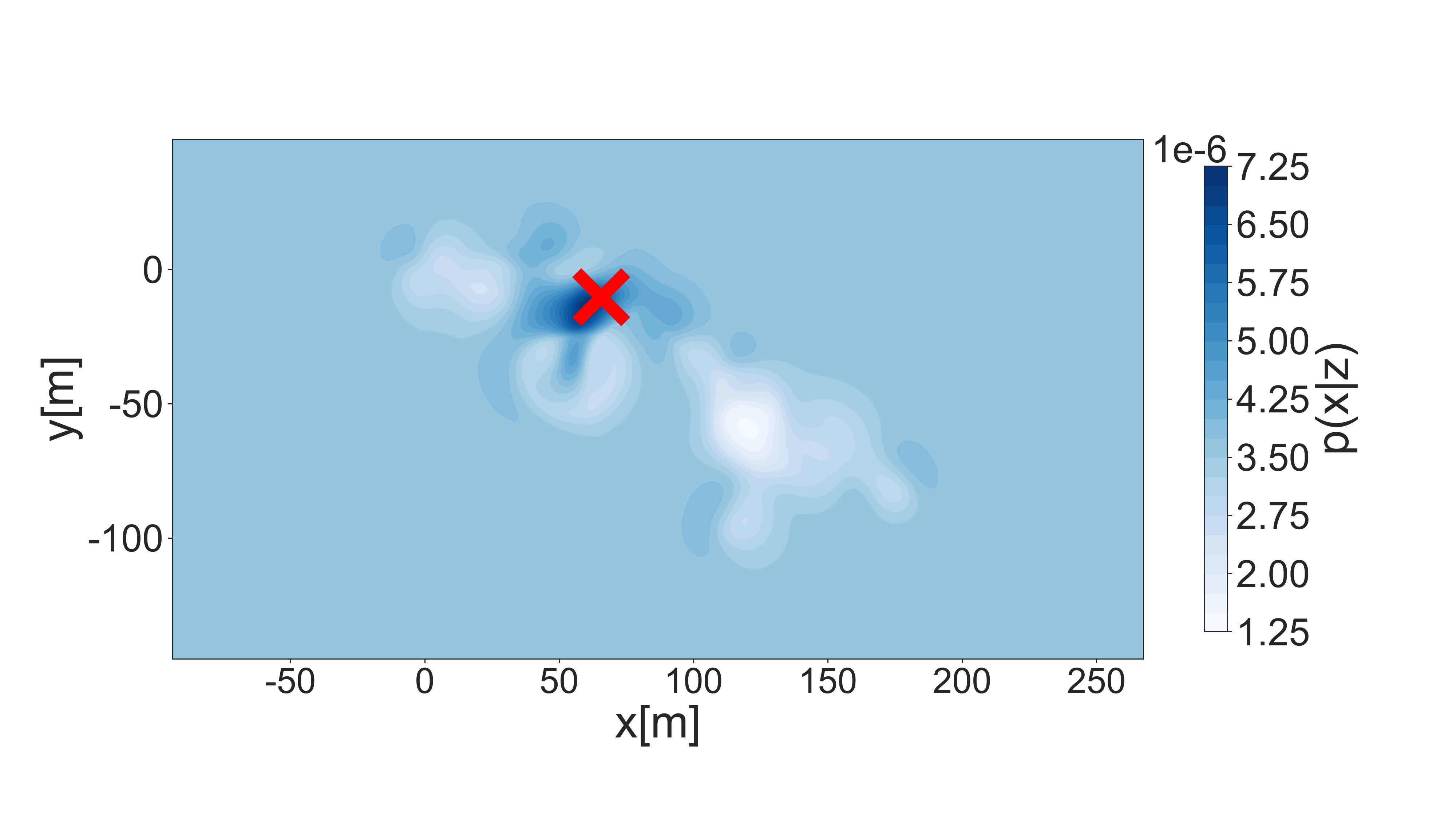}}~
  \subfigure{\includegraphics[width=.22\textwidth]{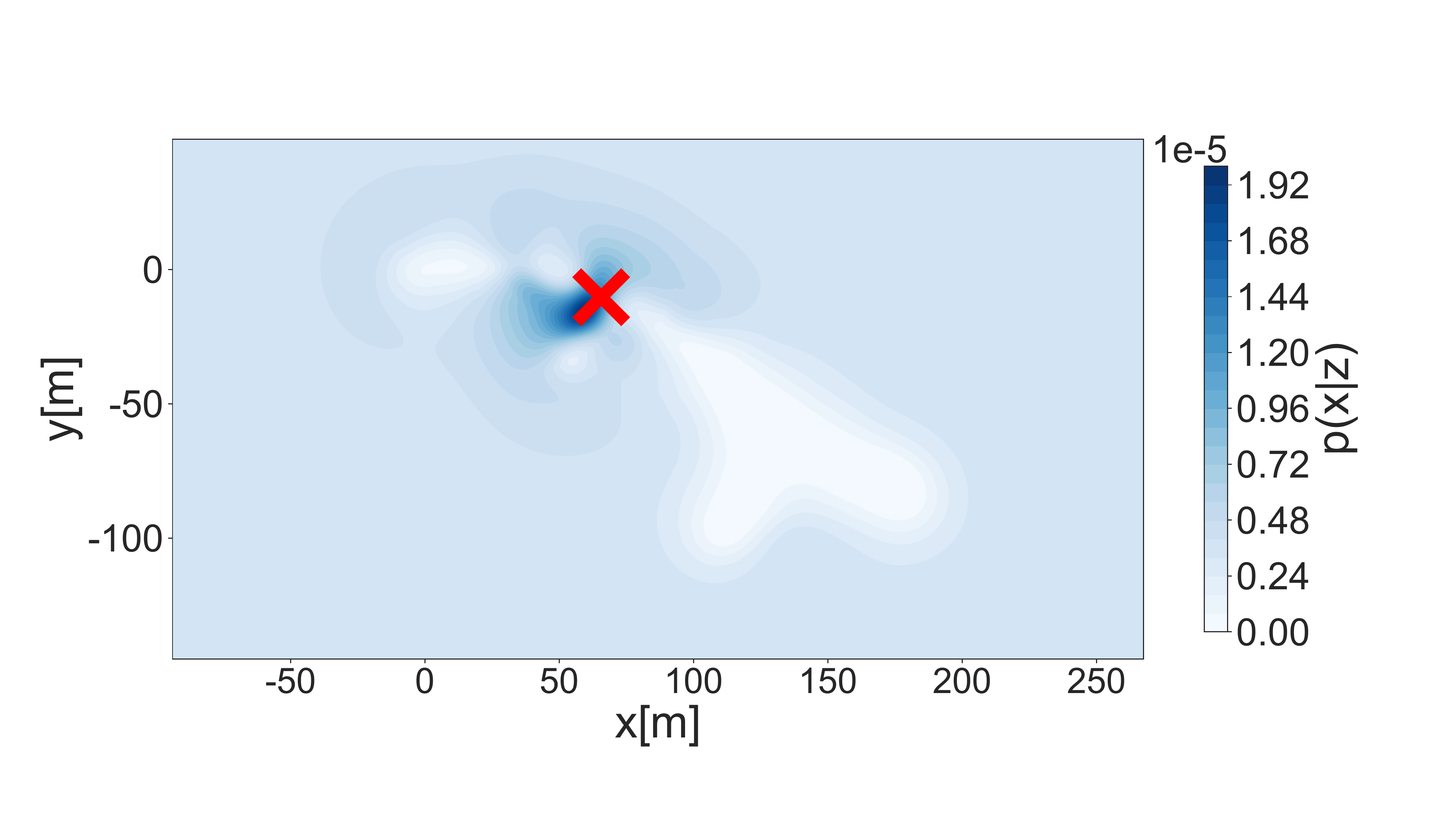}}\newline
  \raisebox{\dimexpr 1.27cm-\height}{(C)}~
  \setcounter{subfigure}{0}
  \hspace{-0.36cm}~
  \subfigure[Input space (91d)]{\includegraphics[width=.22\textwidth]{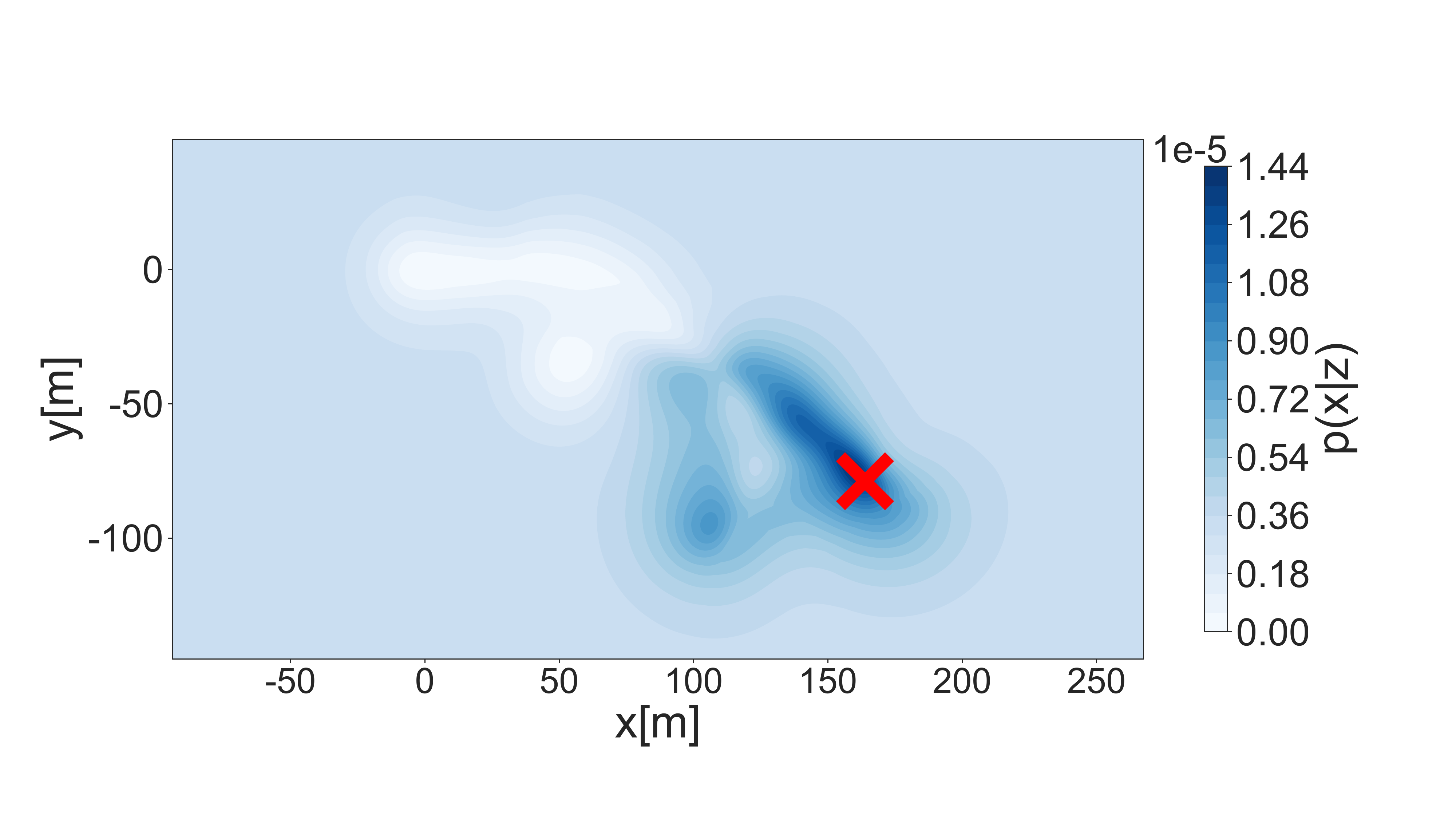}}~ 
  \subfigure[PCA (30d)]{\includegraphics[width=.22\textwidth]{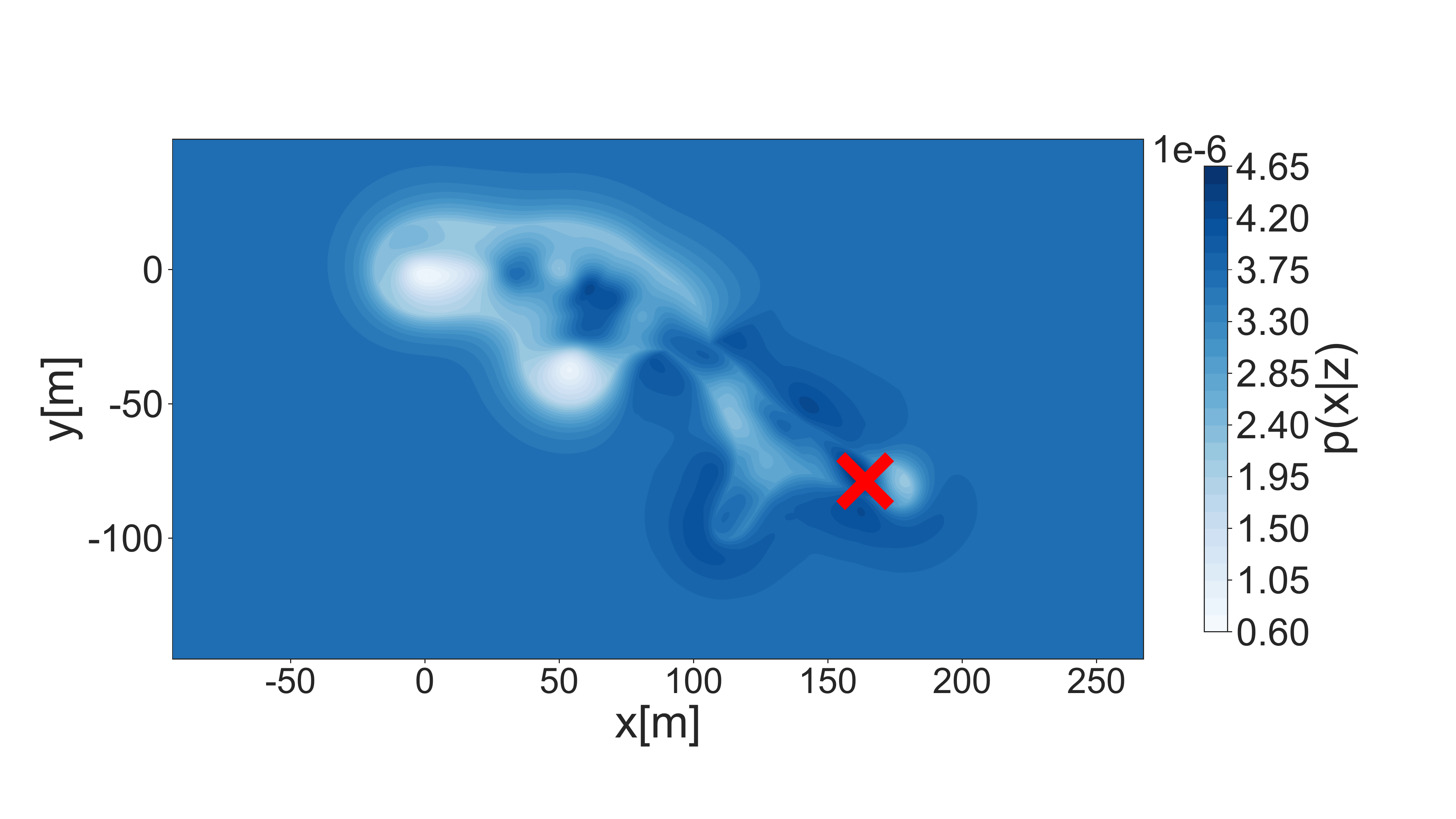}}~ 
  \subfigure[Sparse AE (10d)]{\includegraphics[width=.22\textwidth]{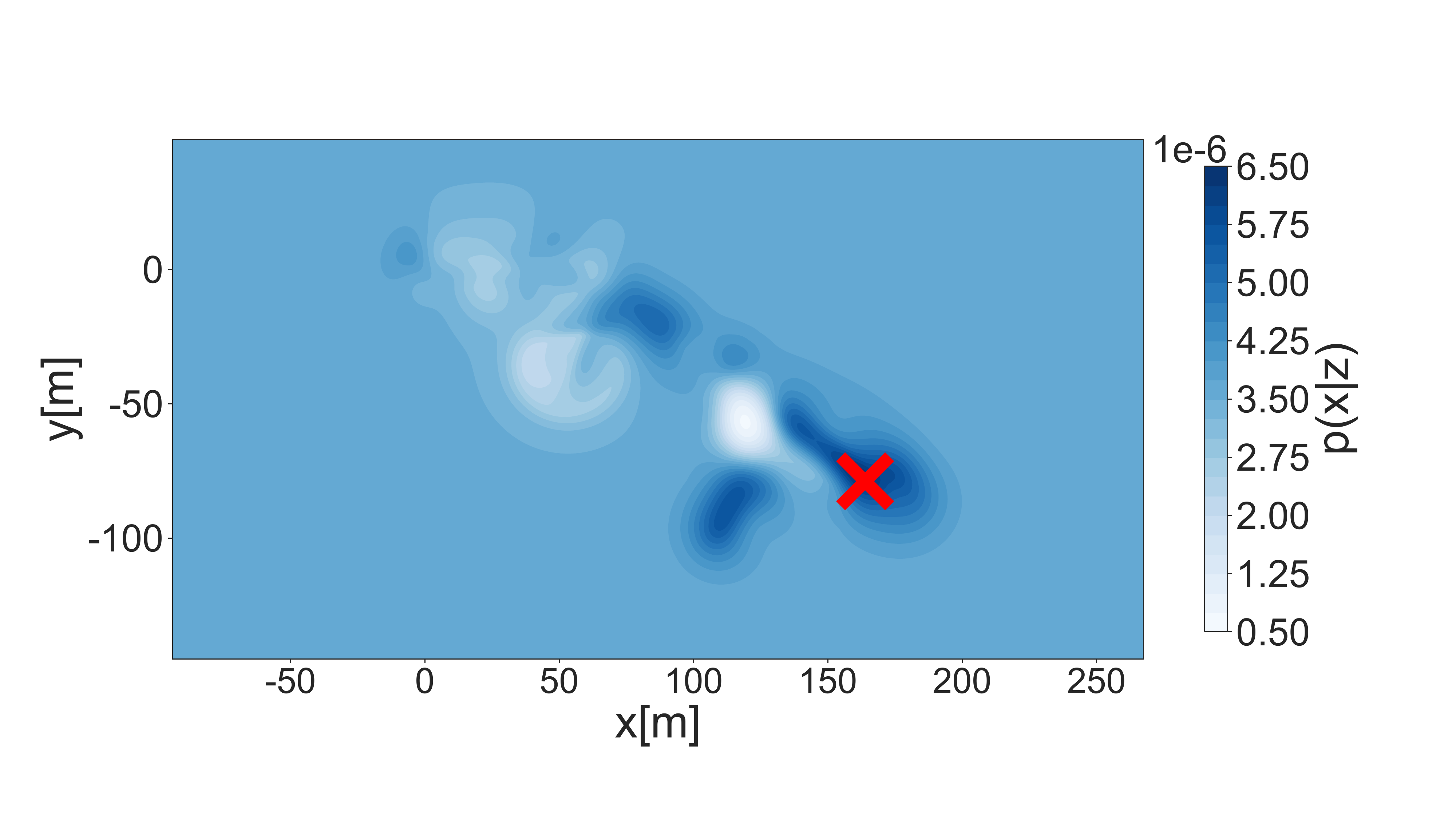}}~
  \subfigure[Our approach (10d)]{\includegraphics[width=.22\textwidth]{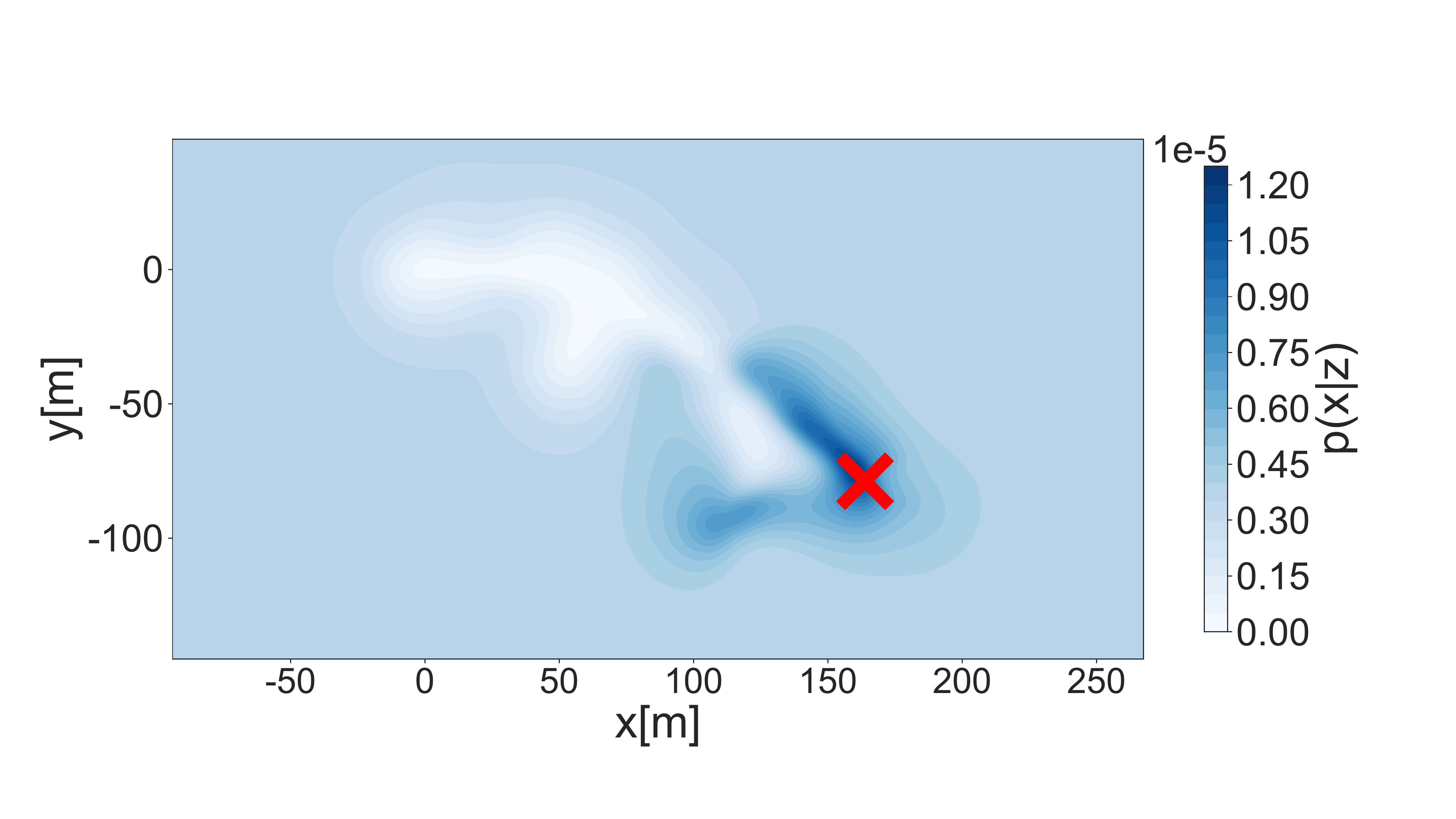}}
  \caption{Example of likelihood functions using the full input space, principal component analisys, sparse autoencoder and our distance invariant autoencoder. True location is marked with a red X, dark blue shades represent areas with high probability density and lighter shades lower density. An ideal posterior would have most of its probability density concentrated around the true location X.\label{fig:example}} 
\end{figure*} 

\subsection{Testing environment and dataset}
For the evaluation of our proposed approach, we surveyed an outdoor environment the innovation corridor at Haneda Innovation City, a large-scale commercial complex attached to Haneda Airpot (Japan). Figure~\ref{fig:hic_map} shows the 2D occupancy map of the environment and the path taken while surveying. All tests were performed using a mobile robot with a mounted router, which was modified to continuously acquire signal strength measurements. All data (signal strengths, odometry, and laser rangefinder measurements) was recorded with timestamps in a rosbag and has been made available online\footnote{\href{https://jinko.ir.utsunomiya-u.ac.jp/data}{\url{https://jinko.ir.utsunomiya-u.ac.jp/data}}}. Two runs were performed with the robot, one to collect the training dataset and another for testing our approach. The robot was operated continuously at walking speed and signal. Signal strength information was recorded only from beacon frames to guarantee that signals come from access points and not mobile users, and odometry and range data are only used for building occupancy maps and obtaining datasets' ground truth locations for training and testing.  91 different access points were identified in the area.

\subsection{Autoencoder}
For this environment we chose the encoder function to have 60 units in the first layer and 10 units in the output layer, compressing the data from 91 to 10. The decoder function was similarly chosen to have 60 units in the first layer and 91 layers on the output layer.

\subsubsection{Reconstruction}
\begin{figure}[!t] 
\centering 
\subfigure[$\Z$]{\includegraphics[height=.6\columnwidth]{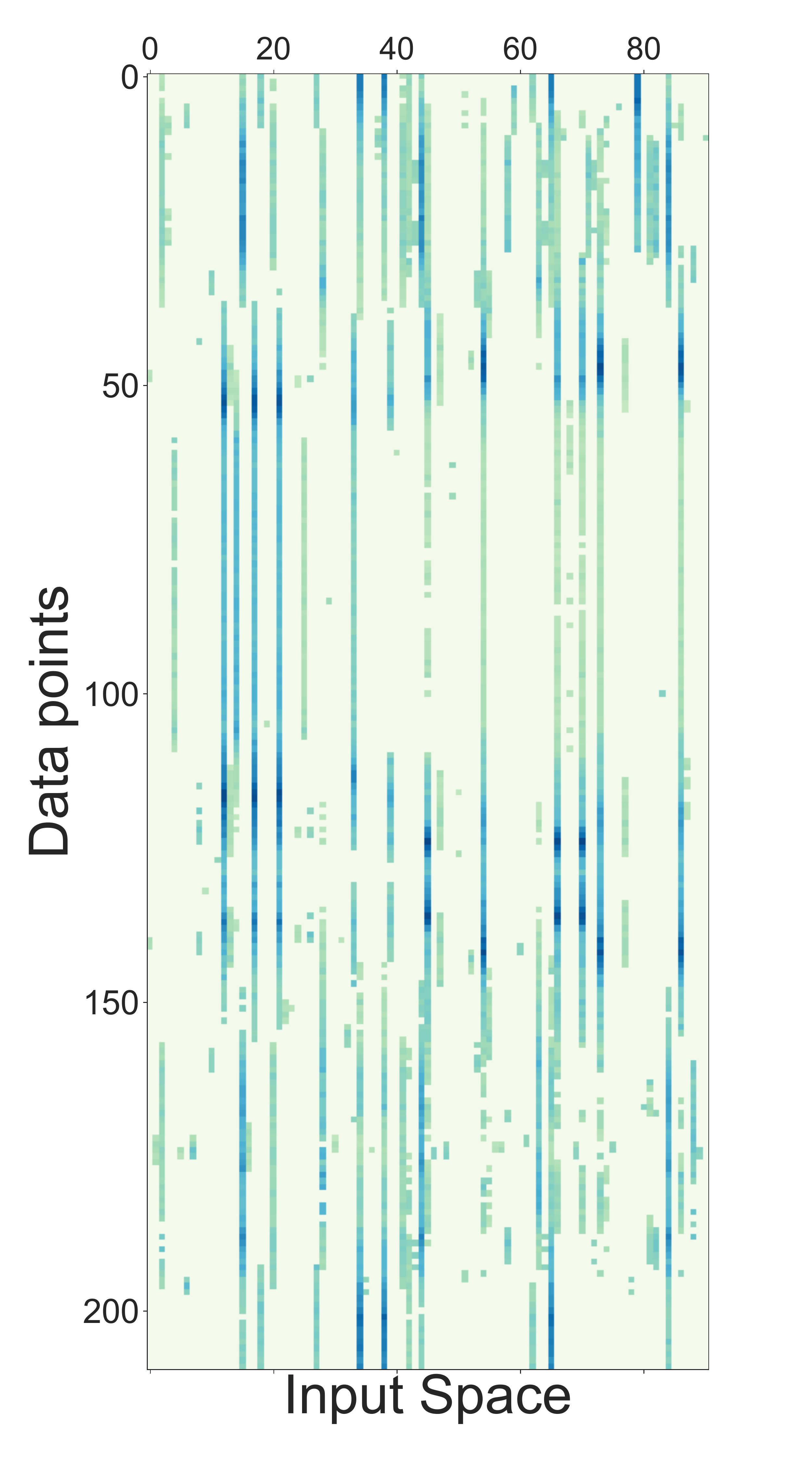}}~ 
\subfigure[$\mathbf{L}$]{\includegraphics[height=.6\columnwidth]{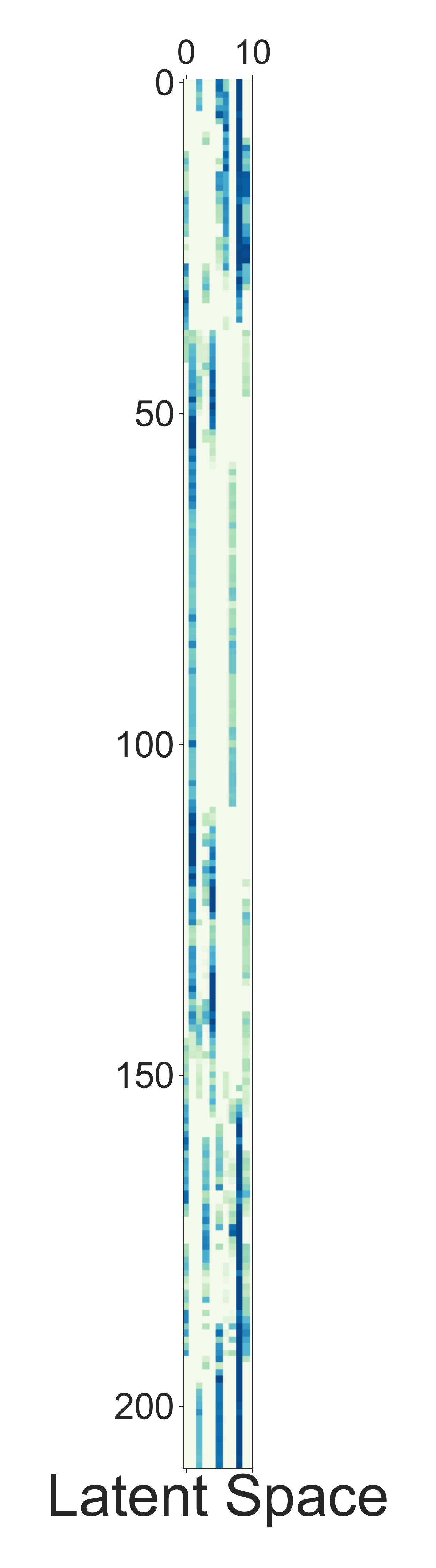}}~ 
\subfigure[$\hat{\Z}$]{\includegraphics[height=.6\columnwidth]{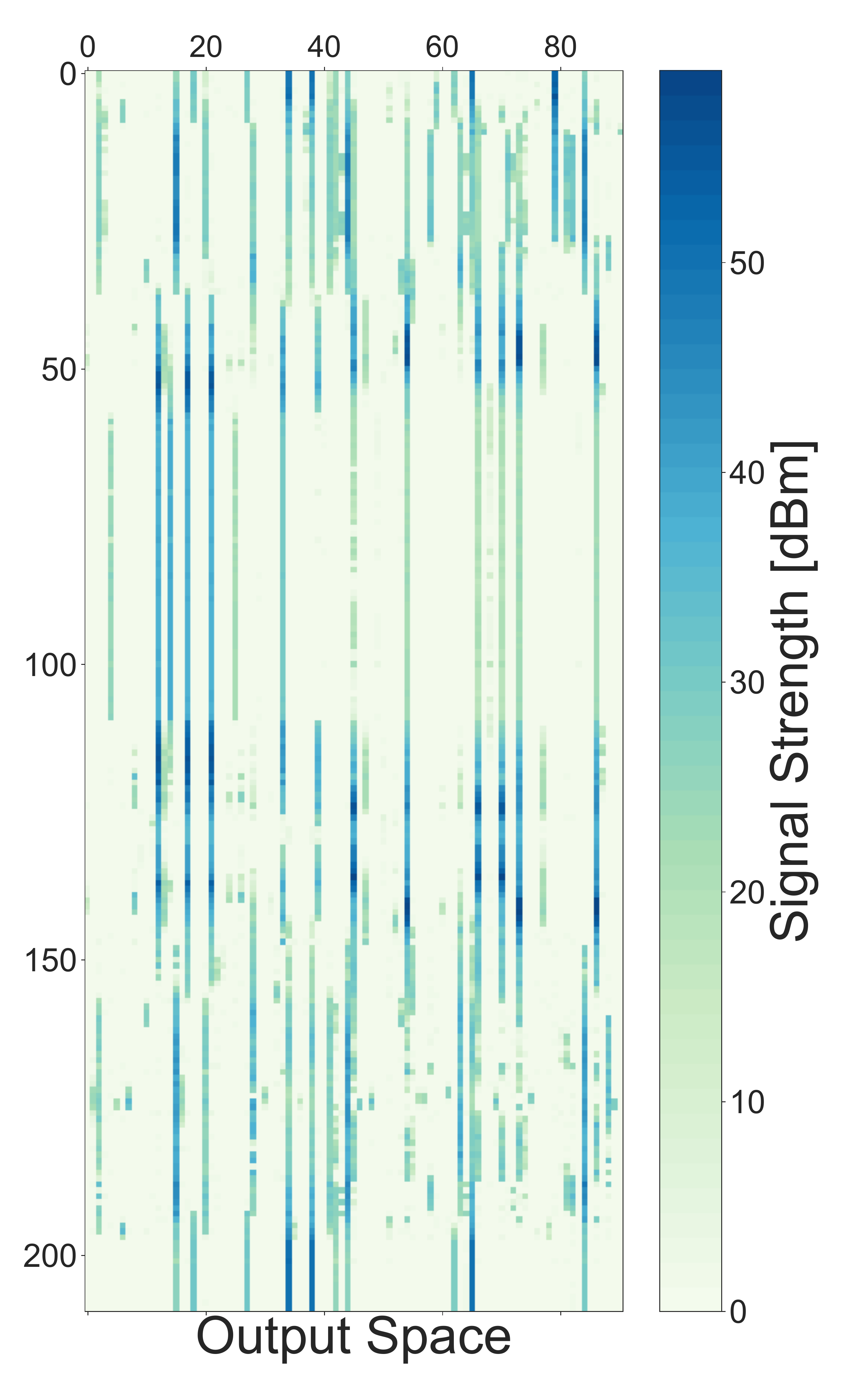}}\newline\vspace{-0.1cm}
\subfigure[RMSE reconstruction errors per access point]{\includegraphics[width=0.8\columnwidth]{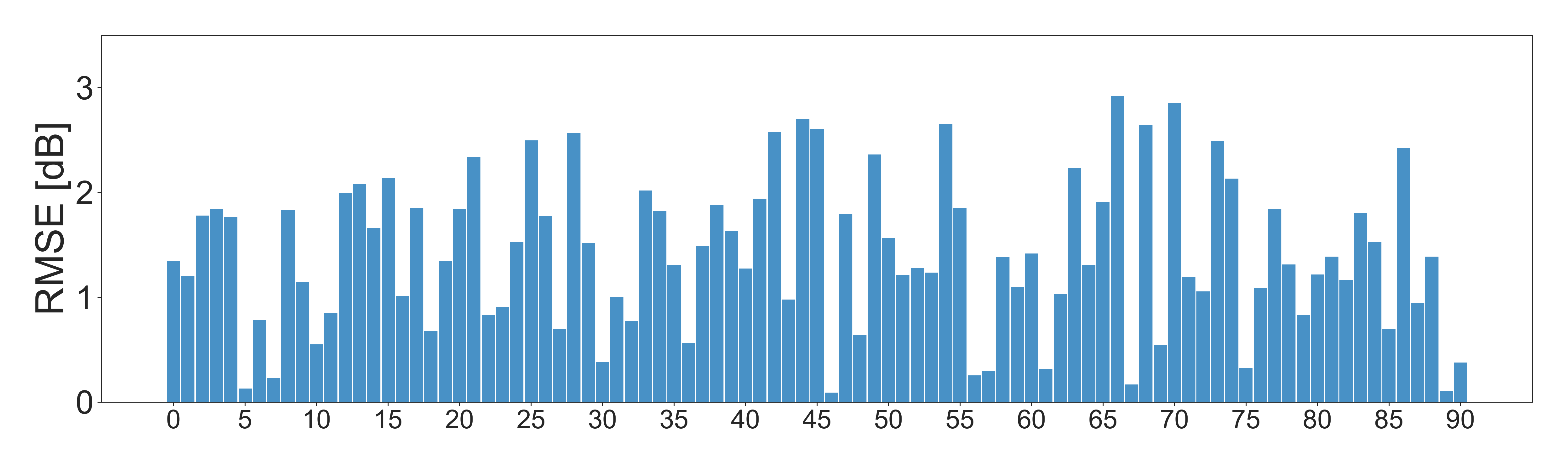}}
\caption{Training data in (a) input, (b) latent and (c) reconstruction spaces\label{fig:reconstruction}} 
\end{figure}

Figure~\ref{fig:reconstruction} shows the (a) training datasets $\Z$, (b) compressed dataset $\mathbf{L}$ and (c) reconstructed dataset $\hat{\Z}$. As it can be observed, not much information was lost during reconstruction with an average root mean squared error of 1.6 dBm overall. Figure~\ref{fig:reconstruction}(d) shows the root mean squared errors per access point, with the maximum errors around 3 dBm.

\subsection{Joint likelihoods from latent spaces}
\subsubsection{Qualitative analysis}
Figure~\ref{fig:example} shows examples of likelihood functions at 3 different locations in the environment. The likelihoods were generated from maps learned using the same GP model described in our approach on different spaces: (a) Original input space (91 dimensions), latent space learned using (b) principal component analysis (30 dimensions), (c) sparse autoencoders - equivalent to using only the reconstruction and regularization losses (10 dimensions), and (d) distance invariant sparse autoencoders (our approach) (10 dimensions)

Considering that ideally, posteriors should have high probability densities for locations near the true location (marked with an X), and lower otherwise, we can observe that for location (A), all generated posteriors are similar and adequate. For locations (B) and (C) the PCA approach fails to generate adequate posteriors although having a much lower reduction than the autoencoders. In general, for all locations, we can observe that sparse autoencoders tend to output less peaked distributions (larger probability densities far from the true location) than the input and our proposed latent space, especially at location (B).

\subsubsection{KL-divergence}
To quantify the quality of the generated maps, we discretized the environment using a grid of fixed size $\mathcal{X}: \{\x_i\}^{i=1:k}$, compute the likelihood for all locations and then the Kullback–Leibler divergence (KL-divergence) between this distribution and an ideal distribution. The ideal distribution we chose was a Gaussian centered on the ground truth with a standard deviation of 10 meters.

\begin{figure}[tb] 
\centering 
\includegraphics[width=\columnwidth]{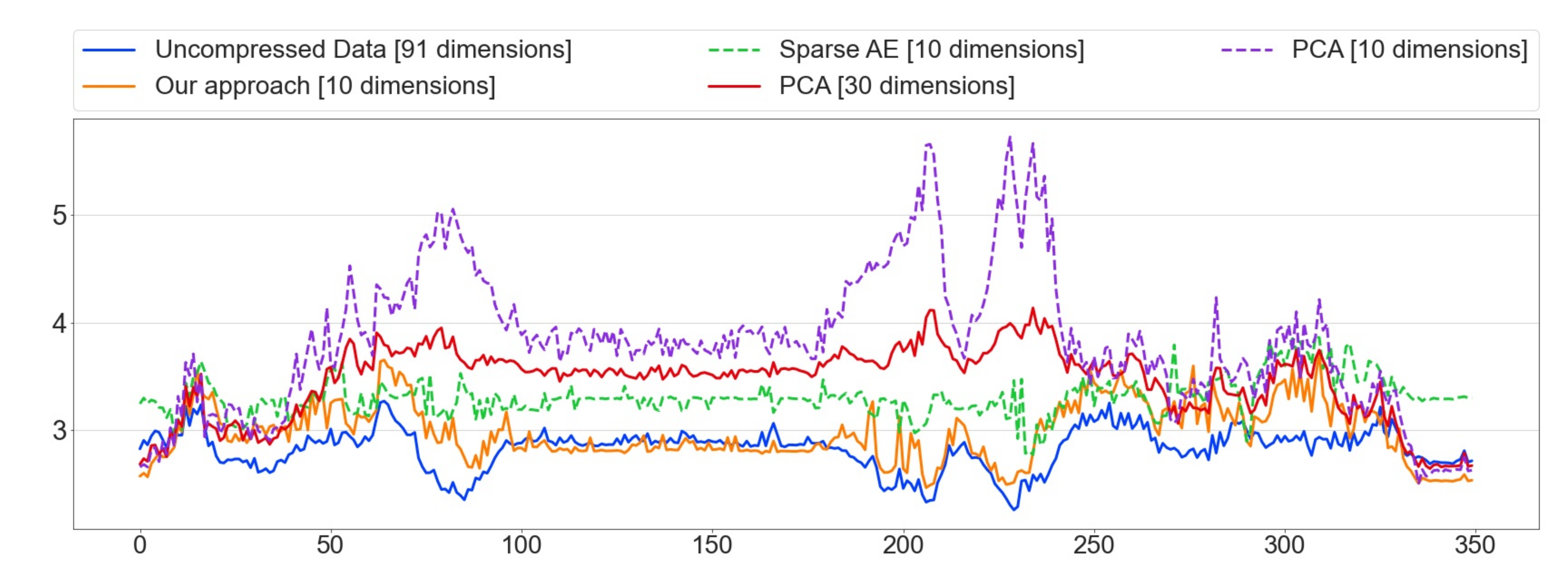} 
\caption{KL-divergence comparing the different approaches at all testing points (lower is better)\label{fig:kl}} 
\end{figure}

Figure~\ref{fig:kl} shows the KL-divergence for the models previously mentioned (with PCA\_30 being a PCA with 30 dimensions and PCA\_10 one with 10 dimensions). 
As expected, overall, the uncompressed model is still performing the best, with an average score of 2.82. Our approach follows second at 2.98, even outperforming the uncompressed model at some testing points. The advantage of our distance invariance measure becomes obvious as our approach consistently outperforms sparse autoencoders that scored on average 3.31, with both models considerably outperforming PCA - 3.46 for the one with 30 dimensions and 3.82 for the one with 10 dimensions.

\section{Conclusions and Future works\label{sec:conclusion}}
In this work, we have proposed a new autoencoder that encourages isometry between input and latent spaces using a new distance invariance loss. 

We tested our approach experimentally in an outdoor where 91 different access points were located successfully compressing this space to 10 dimensions and reconstructing signals from this latent space accurately (average of 1.6 dBm RMSE). Furthermore, using its encoder we were able to learn location-to-signal maps in this compressed space and used them directly to compute the likelihood functions required for Bayes-filters. Compared to PCA and sparse autoencoders our likelihoods were more similar to theoretically ideal distributions, and only slightly worse than those computed without using any compression. Therefore, not only did we reduce the required amount of memory for storing maps over 9 folds, but also similarly reduced the computation required for calculating likelihoods (though there is the added overhead of transforming new signals to the latent space). 

Future research will use these likelihoods as the sensor model of a Monte Carlo Localization algorithm, explore the use of variational autoencoders (probabilistic versions of the autoencoders we used), other rules to further encourage isometry, as well as including data augmentation stages so we can train more robust or deeper networks. 

\bibliographystyle{IEEEtran}
\balance
\bibliography{../MiyagusukuBIB}

\end{document}